\documentclass[12pt]{article}

\ifx\pdfoutput\undefined
\usepackage[dvips,bookmarks]{hyperref}
\else
\usepackage{hyperref}
\fi
\hypersetup{colorlinks=false,bookmarksopen,bookmarksnumbered,citecolor=blue,
   pdfstartview=FitH}

\usepackage{latexsym}

\usepackage{amssymb,amsfonts,amsmath}
\usepackage{graphicx} 
\usepackage{indentfirst}

 \usepackage{bbm}
\usepackage{ulem} 
\usepackage{color}

\parskip=\medskipamount

\newcommand {\cJ}{{\cal J}}

\newcommand {\cN}{{\cal N}}

\def\a{\alpha}

\def\b{\beta}
\def\c{\chi}
\def\d{\delta}

\def\f{\phi}
\def\g{\gamma}
\def\G{\Gamma}

\def\k{\kappa}
\def\l{\lambda}

\def\o{\omega}

\def\r{\rho}
\def\s{\sigma}

\def\D{\Delta}
\def\F{\Phi}
\def\J{\Psi}

\def\S{\Sigma}
\def\U{\Upsilon}

\def\rd{{\rm d}}
\def\ri{{\rm i}}
\def\re{{\rm e}}

\newcommand{\ad}{{\dot{\alpha}}}                           
\newcommand{\bd}{{\dot{\beta}}}                          

\newcommand{\pa}{\partial}                           
\newcommand{\hf}{\frac12}

\newcommand{\vf}{\varphi}

\newcommand{\be}{\begin{equation}}
\newcommand{\ee}{\end{equation}}
\newcommand{\bea}{\begin{eqnarray}}
\newcommand{\eea}{\end{eqnarray}}

\newcommand{\ba}{\begin{array}}
\newcommand{\ea}{\end{array}}

\def\double #1{#1{\hbox{\kern-2pt $#1$}}}

\newcommand{\bsubeq}{\begin{subequations}}
\newcommand{\esubeq}{\end{subequations}}

\begin{document}

\begin{titlepage}

\begin{flushright}
June 2013\\
\end{flushright}
\vspace{5mm}

\begin{center}
{\Large \bf  Self-dual supersymmetric nonlinear sigma models }
\end{center}

\begin{center}

{\large  
S. M. Kuzenko
and I. N. McArthur\footnote{{ian.mcarthur@uwa.edu.au}}
} \\
\vspace{5mm}

\footnotesize{
{\it School of Physics M013, 
The University of Western Australia\\
35 Stirling Highway, Crawley W.A. 6009 Australia }}  
~\\

\vspace{2mm}

\end{center}
\vspace{5mm}

\begin{abstract}
\baselineskip=14pt
In four-dimensional ${\cal N}=1$ Minkowski superspace,
general nonlinear $\sigma$-models with four-dimensional target spaces 
may be realised in term of CCL (chiral \& complex linear) dynamical variables 
which consist of a chiral scalar, a complex linear scalar and their conjugate superfields.
Here we introduce CCL $\s$-models that 
are invariant under U(1) ``duality rotations'' exchanging the dynamical variables and their equations of motion. The Lagrangians of such  
$\sigma$-models prove to obey a partial differential equation that is analogous
to the self-duality equation obeyed by
U(1) duality invariant models for nonlinear electrodynamics. 
These $\sigma$-models are 
self-dual under a Legendre transformation that 
simultaneously dualises (i) the chiral multiplet into a complex linear one;
and (ii) the complex linear multiplet into a chiral one.
Any CCL $\sigma$-model possesses a dual formulation given in terms of two chiral multiplets. 
The U(1) duality invariance of the CCL $\sigma$-model proves to be equivalent, 
in the dual chiral  formulation, to a manifest U(1) invariance rotating the two chiral scalars.
Since the target space has a holomorphic Killing vector, 
the $\sigma$-model possesses a third formulation realised in terms of 
a chiral multiplet and a tensor multiplet.

The family of U(1) duality invariant CCL $\sigma$-models 
includes a subset of ${\cal N}=2$ supersymmetric theories. 
Their target spaces are hyper K\"ahler manifolds with a non-zero Killing vector
field. In the case that the Killing vector field is triholomorphic, 
the $\sigma$-model admits a dual formulation in terms of 
a self-interacting off-shell ${\cal N}=2$ tensor multiplet. 

We also identify a subset of CCL $\sigma$-models which are in a one-to-one correspondence 
with the U(1) duality invariant  models for nonlinear electrodynamics. 
The target space isometry group for these sigma models contains a subgroup
$\rm  U(1) \times U(1)$.
\end{abstract}
\vspace{1cm}

\vfill
\end{titlepage}

\tableofcontents{}
\vspace{1cm}
\bigskip\hrule

\section{Introduction}
\setcounter{equation}{0}

Within the framework of four-dimensional $\cN=2$ Poincar\'e supersymmetry, 
the most general nonlinear $\s$-model can be formulated using {\it off-shell} polar 
supermultiplets \cite{LR88} (see \cite{LR2008,K-lectures} for reviews). From the point of view 
of $\cN=1$ supersymmetry, the $\cN=2$ polar supermultiplet is equivalent to an infinite set 
of $\cN=1$ superfields that naturally split into the categories physical and auxiliary.
The physical superfields consist of a chiral scalar $\Phi$ 
and a complex linear scalar $\S$ 
constrained by 
\bea
\bar{D}_{\ad} \Phi= 0~, \qquad  \bar{D}^2  \S = 0~,
\label{constraints}
\eea
as well as their conjugates $\bar \F $ and $\bar \S$. 
The auxiliary superfields consist of {\it unconstrained} complex 
scalars $\U_n$ and their conjugates $\bar \U_n$, where $n=2,3, \dots$ 
Formulated in $\cN=1$ superspace, the $\s$-model action  
has the form\footnote{Capital Latin letters from 
the middle of the alphabet are used to denote the target-space indices.}
\bea
S= \int \rd^4x \rd^4 \theta \, {\mathfrak  L}(\Phi^I, \bar{\Phi}^{\bar J}, \S^I, \bar{\S}^{\bar J}, \U^I_n , \bar \U^{\bar J}_n)~, 
\label{1.1}
\eea 
for some Lagrangian $\mathfrak  L$. Since the auxiliary superfields 
$\U^I_n$ and $\bar \U^{\bar J}_n$ are unconstrained and appear in the action without derivatives, 
they may be in principle integrated out (using the techniques developed over the last 15 years 
\cite{GK1,GK2,AN,AKL1,AKL2,KN,K-duality,K-comments,BKLT-M}). 
Then the $\s$-model action turns into 
\bea
S= \int \rd^4x \rd^4 \theta \, L(\Phi^I, \bar{\Phi}^{\bar J}, \S^I, \bar{\S}^{\bar J})~.
\label{1.2}
\eea  
Any $\cN=2$ supersymmetric nonlinear $\s$-model can be recast in  the $\cN=1$ superfield  form \eqref{1.2}, and for this reason such  $\cN=1$ nonlinear $\s$-models have been the focus of much attention.
The Lagrangian $L$ in \eqref{1.2}
must obey some nontrivial conditions in order for the action to be 
invariant under {\it on-shell} $\cN=2$ supersymmetry transformations,
 see \cite{K-comments,BKLT-M} for more details. In this note we provide 
 an alternative motivation for study of $\cN=1$ supersymmetric nonlinear $\s$-models involving chiral and complex linear (CCL) superfields -- i.e. of the form \eqref{1.2}.
 In what follows, the  supersymmetric theories \eqref{1.2} will be often referred to as   
 CCL $\s$-models.   
 It should be mentioned that  such nonlinear $\s$-models were discussed for the first time  
long ago by Deo and Gates \cite{DG}.

Chiral and complex linear superfields are known 
to provide dual off-shell descriptions of
the massless scalar supermultiplet \cite{Zumino80,GS,GGRS}.
By performing a special superfield Legendre transformation, which is reviewed in section \ref{section4}, 
any supersymmetric theory that is realised in terms of 
a complex linear scalar $\S$ and its conjugate $\bar \S$ 
has a dual formulation in terms of  a chiral scalar $\J$, $\bar D_\ad \J=0$, 
and its conjugate $\bar \J$.
When applied to the $\s$-model \eqref{1.2}, 
this leads to a purely chiral formulation\footnote{It is well-known 
that any $\cN=1$ or $\cN=2$ supersymmetric $\s$-model can be formulated in terms of $\cN=1$
 chiral scalars and their conjugates \cite{Zumino,HKLR}.}
\bea
S= \int \rd^4x \rd^4 \theta \, K(\Phi^I, \bar{\Phi}^{\bar J}, \J_I  , \bar{\J}_{\bar J})~,
\label{1.3}
\eea  
where $K$ is the K\"ahler potential for a target space.
On the other hand, the inverse Legendre transformation allows us 
to dualise chiral multiplets into complex
linear ones. Both the Legendre transformation and its inverse may be 
applied simultaneously to the variables $\Phi$ and $\S$ in (\ref{1.2}). 
One can construct $\s$-models \eqref{1.2} that are self-dual under 
this simultaneous Legendre transformation, in that the dual Lagrangian is equivalent to the original \cite{KLvU}.\footnote{One may also define self-dual $\s$-models 
with manifest $\cN=2$ supersymmetry \cite{KLvU} using the polar-polar duality \cite{GK1}
(see also \cite{LR2008}).} 

Self-duality under a Legendre transformation is 
an example of a discrete duality symmetry. In this note we will be interested 
in nonlinear $\s$-models 
\eqref{1.2} which possess a continuous U(1) duality invariance. 
It turns out that such $\s$-models display a remarkable similarity with the U(1) 
duality invariant models for nonlinear electrodynamics 
\cite{GZ1,GR1,GR2,GZ2,GZ3} (see \cite{KT2,AFZ} for reviews).

This paper is organised as follows. In section 2 we present the general theory of U(1) 
duality invariant CCL $\s$-models. In section 3 we identify a subset of CCL 
$\s$-models which are in a one-to-one correspondence 
with the U(1) duality invariant  models for nonlinear electrodynamics. 
Section 4 is devoted to the dual chiral formulation. Dual Born-Infeld type solutions
are considered in section 5. Concluding comments are given in section 6. 
Finally, the Appendix is devoted to the derivation of the duality equation.

\section{Duality invariant sigma models}\label{section2}
\setcounter{equation}{0}

The object of our study is a  supersymmetric nonlinear $\s$-model
 realised in terms of a chiral scalar $\F$, a complex linear  scalar $\S$ 
 and their conjugates.
The action is 
\be
S_{\text{CCL}}= \int \rd^4x \rd^4 \theta \, L(\Phi, \bar{\Phi}, \S, \bar{\S})~.
\label{2.1}
\ee  
The off-shell constraints \eqref{constraints} can always be solved 
in terms of unconstrained complex {\it gauge} superfields as 
$\Phi = \bar{D}^2 \bar U $ and $\S= \bar{D}_{\ad} \bar \r^{\ad}$, 
and therefore the equations of motion 
are
\bea \bar{D}_{\ad} \, \frac{ \partial L}{\partial \S} = 0~, \qquad 
 \bar{D}^2 \, \frac{\partial L}{\partial \Phi} = 0~. 
\eea
We see that the equations of motion have the same functional form as the off-shell 
constraints but with $\F$  and $\S$ interchanged.
As a result, ``duality'' rotations that mix $\Phi$ and 
${ \partial L}/{\partial \S}$,
 and $\S$ 
and ${ \partial L}/{\partial \Phi},$ 
leave the constraints and the equations of motion invariant.\footnote{These are motivated by the duality rotations in vacuum electrodynamics, which mix the Bianchi identities $ \vec{\nabla} \times \vec{E} = 0$ and $\vec{\nabla} \cdot \vec{B} = 0$ with equations of motion $\vec{\nabla} \times \frac{\partial L}{ \partial \vec{B}} = 0$ and $\vec{\nabla} \cdot\frac{\partial L}{ \partial \vec{E}} = 0.$ In this case, the rotations are termed duality rotations because they mix derivatives of the field strength,  $ \partial^{a} F_{ab},$ with derivatives of the  Hodge dual of the field strength, $\partial^a \tilde{F}_{ab}.$ 
We may think of the superfields $\F= \bar{D}^2 \bar U $  and $\S= \bar{D}_{\ad} \bar \r^{\ad}$ as gauge invariant field strengths.}

We consider  the continuous  U(1) duality rotations\footnote{The compact 
U(1) duality transformations may be promoted to non-compact ones
in the presence of additional matter fields, 
in complete analogy with 
nonlinear electrodynamics.}
\begin{subequations}\label{2.3}
\bea
\left(
\begin{array}{c}
\Phi' \\  \frac{\partial L'(X')}{\partial \S'}
\end{array}
\right)
&=& \left(
\begin{array}{cc}
\cos \l & \, \, \sin \l   \\  - \sin \l & \, \, \cos \l 
\end{array}
\right) \, \, \left(
\begin{array}{c}
\Phi\\  \frac{\partial L(X)}{\partial \S}
\end{array}
\right)  \label{2.3a}
\eea
and
\bea
\left(
\begin{array}{c}
\S' \\  \frac{\partial L'(X')}{\partial \Phi'}
\end{array}
\right)
&=& \left(
\begin{array}{cc}
\cos \k & \, \, \sin \k   \\  - \sin \k & \, \, \cos \k 
\end{array}
\right) \, \, \left(
\begin{array}{c}
\S \\  \frac{\partial L(X)}{\partial \Phi}
\end{array}
\right)~, \label{2.3b}
\eea
where, for notational convenience, the symbol $X$ has been introduced for the argument $\Phi,  \bar{\Phi}, \S,\bar{\S}.$
The left hand sides of these equations define the chiral scalar $\Phi',$ the complex linear scalar $\S'$ and the Lagrangian $L'.$ As shown in the Appendix, there is an integrability condition associated with the existence of the Lagrangian $L'(X')$ defined in this way that forces the condition 
\bea
\k = \l~.
\eea
\end{subequations}

Duality invariance of the theory is the requirement that the Lagrangians
$L'$ and $L$ have the same functional form,
\be
L'(X') = L(X')~.
\label{dualityinv}
\ee
The implications of this condition 
 are derived in the Appendix, mimicking standard constructions in 
 nonlinear electrodynamics (see, for example, the Appendix in \cite{KT2}). 
 They are that   the Lagrangian $ L(X) =L(\Phi,\bar{\Phi},  \S, \bar{\S}) $ 
must obey the differential equation
\be
0 = \Phi \, \S +    \bar{\Phi} \, \bar{\S} 
 +  \frac{\partial L }{\partial \Phi} \, \frac{\partial L }{\partial \S}
+  \frac{\partial L  }{\partial \bar{\Phi}}  \, \frac{\partial L }{\partial \bar{\S}} 
 ~.
\label{SS}
\ee

Note that the condition  \eqref{dualityinv}
is not equivalent to the requirement that the Lagrangian itself 
is invariant under the duality rotations. 
 Given a duality invariant theory, its Lagrangian varies as 
 \be
 \d L(X) \equiv L(X') -L(X)
= 
-2\l\big(  \Phi\,  \S + \bar{\Phi} \, \bar{\S} \big)
~,
 \label{delta L}
 \ee 
 as a consequence of \eqref{SS}. 
 However, there exist two general prescriptions to construct  duality invariant observables 
 starting from $L$.
 
Firstly, making use of eq. (\ref{SS}), one finds that 
\be
\d \left( L - \frac12 \, \Phi \, \frac{\partial L }{\partial \Phi} 
- \frac12 \, \bar{\Phi} \, \frac{\partial L}{\partial \bar{\Phi}} 
- \frac12 \, \S \, \frac{\partial L}{\partial \S} 
- \frac12 \, \bar{\S} \, \frac{\partial L }{\partial \bar{\S}} \right)  = 0~,
\ee
and so the expression in parentheses is invariant under 
arbitrary duality rotations.
For a finite duality rotation  (\ref{2.3}), this means that 
\bea
& & L(X) - \frac12\,  \Phi\,  \frac{\partial L(X)}{\partial \Phi} 
- \frac12 \, \bar{\Phi} \, \frac{\partial L(X)}{\partial \bar{\Phi}} 
- \frac12 \, \S \,\frac{\partial L(X)}{\partial \S} 
- \frac12 \, \bar{\S} \, \frac{\partial L(X)}{\partial \bar{\S}} \nonumber \\
&= & L(X') - \frac12 \, \Phi'  \, \frac{\partial L(X')}{\partial \Phi'} 
- \frac12 \, \bar{\Phi}' \, \frac{\partial L(X')}{\partial \bar{\Phi}'} 
- \frac12  \, \S' \, \frac{\partial L(X')}{\partial \S'} 
- \frac12 \, \bar{\S}' \, \frac{\partial L(X')}{\partial \bar{\S}'}~.
\label{finite}
\eea

Secondly, if the Lagrangian depends on a duality invariant parameter $g$, 
that is $L = L(X;g)$, then the observable $\pa L(X;g)/\pa g$ is duality invariant. 
This follows from the fact that the right-hand side of \eqref{delta L} is independent
of $g$. A nontrivial application of this property is that the supercurrent multiplet
of any duality invariant $\s$-model is duality invariant. 

Even though the U(1) duality invariance is not a symmetry of the Lagrangian, 
it leads to the existence of a conserved current $j^a$ such that  $\pa_a j^a=0$.
It is proportional to the component $(\s^a)^{\bd\g}[D_\g, \bar D_\bd ]\cJ|$ 
of the superfield
\bea
\cJ :=\ri \Big(  \Phi \, \S 
 +  \frac{\partial L }{\partial \Phi} \, \frac{\partial L }{\partial \S}\Big) 
=-\ri \Big( \bar  \Phi \, \bar \S 
 +  \frac{\partial L }{\partial \bar  \Phi} \, \frac{\partial L }{\partial \bar\S}\Big)  ~,
 \label{current}
 \eea
 which is real, as a consequence of \eqref{SS}, and obeys the conservation equation 
 \bea
 \bar D^2 \cJ= D^2 \cJ=0
  \label{current2}
 \eea
 on the mass shell. This is similar to the Gaillard-Zumino conserved current 
 in duality invariant electrodynamics \cite{GZ1}.

 An important property of any solution of \eqref{SS} is self-duality 
 under a Legendre transformation which dualises the (anti)chiral variables
 $\F$ and $\bar \F$ into a complex linear superfield $\G$ and its conjugate $\bar \G$,
 and also dualises the complex linear scalar $\S$ and its conjugate 
 $\bar \S$ into a chiral scalar $\J$ and its conjugate $\bar \J$. Before discussing 
 self-duality, let us first  recall the definition of a dual formulation for the 
most general $\s$-model \eqref{2.1} following  \cite{KLvU}.

Starting from \eqref{2.1}, we introduce a first-order action 
\bea
S_{\text{first-order}}= \int \rd^4x \rd^4 \theta \, 
\Big\{ L(X) + \Psi \, \S + \bar{\Psi} \, \bar{\S} - \G \, \Phi  - \bar{\G} \,\bar{\Phi}
\Big\} ~,
\label{f-o}
\eea  
where again, for notational convenience,  the symbol $X$ denotes the set of 
scalars $\Phi,  \bar{\Phi}, \Sigma,\bar{\S}$. Unlike the original action 
\eqref{2.1}, here $\Phi$ and $\S$ and taken 
to be unconstrained complex superfields.  
The newly introduced  superfields $\J$  and $\G$ are chosen to be chiral 
and complex linear respectively, 
\bea
\bar{D}_{\ad} \J= 0~, \qquad  \bar{D}^2  \G = 0~.
\label{constraints2}
\eea
These superfields play the role of Lagrange multipliers.
The dynamical system \eqref{f-o} is equivalent to the original $\s$-model 
\eqref{2.1}. Indeed, the equations of motion for $\J$ and $\G$
enforce  $\S$ and $\F$ to be complex linear and chiral respectively. 
As a result, the four terms with the Lagrange multipliers on the right of  \eqref{f-o} 
drop out and the action turns into \eqref{2.1}.
On the other hand, we may again start from $S_{\text{first-order}}$
and integrate out the auxiliary superfields $X$ using the equations 
of motion for  $\F$ and $\S,$
\bea 
\G = \frac{\partial L(X)}{\partial \Phi}~,
\qquad 
\Psi = - \frac{\partial L(X)}{\partial \S}~, 
\label{eom2}
\eea
as well as  the conjugate equations. Under quite general assumptions, 
these equations may be uniquely solved by expressing 
the original variables $X:= \{\Phi,  \bar{\Phi}, \Sigma,\bar{\S}\}$ in terms of 
the {\it dual} ones $X_{\rm D}:= \{ \J,  \bar{\J}, \G,\bar{\G} \}$.
As a result, 
we end up with a dual formulation for the $\s$-model \eqref{2.1}, 
\bea
S_{\rm D}= \int \rd^4x \rd^4 \theta \, L_{\rm D}(\J, \bar{\J}, \G, \bar{\G})~,
\label{2.13}
\eea  
where 
\be
L_{\rm D}(X_{\rm D}) = L(X) + \Psi \, \S + \bar{\Psi} \, \bar{\S} - \G \, \Phi  - \bar{\G} \,\bar{\Phi}~.
\label{2.14}
\ee

In the derivation of \eqref{2.13}, the original $\s$-model \eqref{2.1}
was completely arbitrary. Now, we assume that the Lagrangian $L$ in \eqref{2.1} 
is a solution of the duality equation \eqref{SS}.
It is useful to consider a special finite duality rotation (\ref{2.3})
with  $\l = \k = {\pi}/{2} $,
\be
\Phi' = \frac{ \partial L(X)}{\partial \S}~, \quad \S' = \frac{ \partial L(X)}{\partial \Phi}~,
 \quad \frac{\partial L(X')}{\partial \S'} = - \Phi~, 
 \quad \frac{\partial L(X')}{\partial \Phi'} = - \S~.
\ee
The relation  (\ref{finite}) becomes
\bea
& &L(X) -   \Phi\,  \frac{\partial L(X)}{\partial \Phi} -  \S \,\frac{\partial L(X)}{\partial \S} -  \bar{\Phi} \, \frac{\partial L(X)}{\partial \bar{\Phi}} - \bar{\S} \, \frac{\partial L(X)}{\partial \bar{\S}} \nonumber \\
&  =& L \left(\frac{ \partial L(X)}{\partial \S},  \frac{\partial L(X)}{\partial \bar{\S}},  
\frac{ \partial L(X)}{\partial \Phi } , \frac{ \partial L(X)}{\partial \bar{\Phi }}\right)~.
\label{2.16}
\eea
Using the equations of motion (\ref{eom2}) and the definition (\ref{2.14}),
eq. \eqref{2.16} yields 
\be
L_{\rm D}( X_{\rm D} ) = L( X_{\rm D} )~,
\ee
establishing the self-duality of the $\s$-model under the Legendre transformation. 

\section{Sigma model cousins of nonlinear electrodynamics}\label{section3}
\setcounter{equation}{0}

In this section, we consider a subclass of CCL  $\s$-models \eqref{2.1} of the form 
\be
S= \int \rd^4x \rd^4 \theta \, L(\o, \bar \o)~,
\ee  
where the complex variable $\o$ is defined by
\be 
\omega = \bar{\Phi} \Phi  - \bar{\S} \S  + \ri \, (\Phi  \S + \bar{\Phi} \bar{\S}) = (\Phi  + \ri \bar{\S}) \, (\bar{\Phi}  + \ri \S) ~.
\label{omega}
\ee
For such $\s$-models, the condition for duality invariance, eq. \eqref{SS},  
can be recast in the form
\be 
0 = \omega - \bar{\o} - 4 \, \omega \left( \frac{\partial L}{\partial \omega} \right)^2  
+ 4 \, \bar{\o} \left( \frac{\partial L}{\partial \bar{\o}} \right)^2~ .
\label{universal}
\ee
This is equivalent to the equation for duality invariance in 
nonlinear electrodynamics (see, e.g., \cite{KT2} for a review)
formulated in terms of the self-dual combination\footnote{Our definition of $\o$, \eqref{2.7}, 
differs in sign from that used in \cite{KT2}. }
\bea
{ \o} = -F_{\a \b} F^{\a \b} = \hf \Big(\vec{E}\cdot \vec{E} - \vec{B} \cdot \vec{B} 
+2 \ri \vec{E} \cdot \vec{B} \Big)= \hf (\vec{E}+\ri  \vec{B})\cdot  (\vec{E}+\ri  \vec{B}) 
\label{2.7}
\eea
and its conjugate $\bar \o$. 
As a result, to any duality invariant solution $L(\o, \bar{\o})$  of the equation (\ref{universal})  in electrodynamics with $\o = -F_{\a \b} F^{\a \b},$ there corresponds a duality invariant supersymmetric 
nonlinear $\s$-model with Lagrangian $L(\o, \bar{\o})$ in which  $\o$ is the superfield combination \eqref{omega}.

The trivial solution of the duality invariance condition (\ref{universal}) is  
\be
L = \frac12 ( \omega + \bar{\omega} )~,
\ee
which in electrodynamics is Maxwell's Lagrangian
\be L = \hf( \vec{E}\cdot \vec{E} - \vec{B} \cdot \vec{B})~,
\ee 
and in the $\s$-model case is 
\be
L = \Phi \bar{\Phi} - \S \bar{\S} ~.
\label{trivial}
\ee

In electrodynamics, a famous nonlinear solution to the duality invariance condition (\ref{universal}) is the Born-Infeld Lagrangian
\be 
L_{\rm BI} = \frac{1}{g} \, (1 - \D^{\frac12}) 
\label{BI}
\ee
with 
\be 
\D = 1 - g  ( \omega + \bar{\omega} ) + g^2 \frac14 (\omega - \bar{\omega})^2~,
 \label{Delta}
\ee
where $g$ is a coupling constant. The corresponding duality invariant supersymmetric nonlinear 
$\s$-model is defined by the Lagrangian (\ref{BI}) with
\be
\D = 1 - 2 g ( \bar{\Phi} \Phi - \bar{\S} \S) - g^2 (\Phi \S + \bar{\Phi} \bar{\S})^2~.
\ee

Of course, not all supersymmetric Lagrangians $ L(\Phi, \bar{\Phi}, \S, \bar{\S})$ satisfying the condition (\ref{SS}) for duality invariance can be expressed in the form $L( \o ,\bar{\o})$. An example is a  
Born-Infeld variant with $\D$ in (\ref{Delta}) replaced by
\be \widetilde{\D} = 1 - 2 g ( \bar{\Phi} \Phi - \bar{\S} \S) 
- 4  g^2\bar{\Phi} \Phi   \bar{\S} \S \label{Delta1}~.
\ee

For those duality invariant Lagrangians that {\it can} be expressed 
in the form $L(\o, \bar{\o}),$ the Lagrangian 
proves to obey two additional differential equations:
\begin{subequations} \label{3.12}
\bea
0&=&  \frac{\partial L}{\partial \Phi} \Phi 
-  \frac{\partial L}{\partial \bar{\Phi}}  \bar{\Phi }
-  \frac{\partial L}{\partial \S} \S
+ \frac{\partial L}{\partial \bar{\S}}  \bar{\S } ~,
\label{3.12a} \\
0 &=&  \frac{\partial L}{\partial \Phi} \bar{\S} 
-  \frac{\partial L}{\partial \bar{ \S}}  \Phi 
-  \frac{\partial L}{\partial \bar{\Phi}} \S
+  \frac{\partial L}{\partial \S }   \bar{\Phi}   ~.
\label{scale} 
\eea
\end{subequations}
The former equation is the condition for  invariance under U(1) transformations
\be
\delta \Phi = \ri  \l  \,\Phi~, \quad \d {\S} = - \ri  \l  \, \S~, \qquad \l \in {\mathbb R}
\ee
which leave $\o$ invariant. 
Equation \eqref{scale} it also expresses the invariance 
of $\o$ under certain linear transformations:
$\delta \Phi = \ri  \l \, \bar{\S}$ and $ \d \bar{\S} = - \ri  \l \, \Phi$.
However such a  transformation is  not  a symmetry of the theory 
under consideration, since it mixes a chiral scalar superfield with a complex linear superfield, and therefore does not respect the off-shell constraints.

\section{Chiral formulation}
\label{section4}
\setcounter{equation}{0}

As discussed in section 1, 
any nonlinear $\s$-model of the form (\ref{2.1}) 
has a purely chiral formulation which is obtained by performing a 
superfield Legendre transformation that dualises the complex linear 
superfield $\S$ and its conjugate $\bar \S$ into a chiral scalar $\J$ 
and its conjugate $\bar \J$. 
It is worth recalling its derivation. 
Starting from the $\s$-model \eqref{2.1}, we introduce a first-order action
\be
S_{\text{first-order}}= \int \rd^4x \rd^4 \theta \, \Big\{ L(\Phi, \bar{\Phi}, \S, \bar{\S})
 + \Psi \S + \bar{\Psi} \bar{\S} \Big\}~,
\label{4.1}
\ee
where $\S$ is taken to be an unconstrained complex superfield,
while the Lagrange multiplier $\J$ is chosen to be chiral, 
\bea
\bar D_\ad \J =0~.
\eea
The original CCL $\s$-model \eqref{2.1} is obtained from \eqref{4.1} by integrating out 
the Lagrange multipliers $\J$ and $\bar \J$. Instead, we can integrate out 
the auxiliary superfield $\S$ and its conjugate $\bar \S$ using the  corresponding 
equation of motion 
\be
 \frac{\partial L}{\partial \S } + \Psi =0 
\label{eom}
\ee
and the  conjugate equation. This leads to the chiral formulation
\bea
S_{\text{chiral}}= \int \rd^4x \rd^4 \theta \, K(\Phi, \bar{\Phi}, \Psi, \bar{\Psi}) ~,
\label{4.4}
\eea
where the corresponding Lagrangian $K$ is the Legendre transform of $L$,
\be
K(\Phi, \bar{\Phi}, \Psi, \bar{\Psi}) 
= L(\Phi, \bar{\Phi}, \S, \bar{\S}) + \Psi \S + \bar{\Psi} \bar{\S}~.
\label{LD}
\ee
The dual Lagrangian $K(\Phi, \bar{\Phi}, \Psi, \bar{\Psi})$ 
is  the K\"ahler potential for a K\"ahler target space. 

If the original Lagrangian $ L(\Phi, \bar{\Phi}, \S, \bar{\S})$ obeys the duality equation 
\eqref{SS},  then the dual chiral action \eqref{4.4}
possesses a manifest U(1) symmetry.\footnote{A direct analogy exists 
with the case of duality invariant systems of $(p - 1)$ forms and $(d-p-1)$ forms 
in $d$ space-time dimensions, as discussed in section 8 of \cite{KT2}.} 
This follows from  (\ref{SS}), in conjunction with standard properties 
of Legendre transformations, including the equation of motion (\ref{eom}) and 
the following relation:\footnote{The 
chiral superfield $\F$  plays the role of a parameter in the context 
of the Legendre transformation described.} 
\be 
\frac{\partial L}{\partial \Phi} = \frac{\partial K}{\partial \Phi}~.
\ee
The resulting dual version of eq. \eqref{SS} is 
\be
0 = \frac{\partial K}{\partial \Psi} \Phi 
+ \frac{\partial K}{\partial \bar{\Psi} } \bar{\Phi} 
-  \frac{\partial K}{\partial \Phi} \Psi
-\frac{\partial K}{\partial \bar{\Phi}}  \bar{\Psi}  ~.
\label{SO2}
\ee
This is the condition for the invariance of 
$K(\Phi, \bar{\Phi}, \Psi, \bar{\Psi})$  under the  holomorphic  U(1) transformations
\bea
\d \Phi = - \l \Psi~, \quad \d \Psi =  \l \Phi~, \qquad \l \in {\mathbb R}~.
\label{4.8}
\eea
We conclude that  the isometry group of the target K\"ahler space 
is nontrivial, since it contains the U(1) subgroup of holomorphic transformations \eqref{4.8}.
This result has several significant implications.

First of all, the holomorphic U(1) symmetry \eqref{4.8} leads to the existence of 
a conserved U(1) current contained  
in the real superfield
\bea
\cJ:= \ri \Big(  \frac{\partial K }{\partial \J} \F -  \frac{\partial K }{\partial \Phi} \J\Big) =\bar \cJ~,
\label{current3}
\eea
which obeys eq. \eqref{current2} on-shell. One can read off the expression \eqref{current3}
from \eqref{current} by using the standard properties of the Legendre transformation. 

Secondly, consider constructing the most general U(1) invariant $\s$-model.
It  is described by a K\"ahler potential 
\bea
K(\Phi, \bar{\Phi}, \Psi, \bar{\Psi})=
{\mathfrak  F} ( \F^2 + \J^2,  \bar \F^2 + \bar \J^2, 
\bar{\Phi} \Phi + \bar{\Psi} \Psi )~,
\eea
where ${\mathfrak  F} (z, \bar z, w)$ is an arbitrary real function of three real variables. 
Performing the inverse Legendre transformation, 
we end up with the most general duality invariant $\s$-model \eqref{2.1}.
Therefore, the chiral formulation provides a generating formalism for 
U(1) duality invariant CCL $\s$-models. 

Thirdly, it is well known \cite{LR83,HoweKLR} that, given a K\"ahler manifold 
with a holomorphic Killing vector field, the  $\cN=1$ supersymmetric $\s$-model 
associated with this target space may be formulated in terms of a single real 
linear superfield, $G=\bar G$, constrained by $\bar D^2 G = 0$ 
(which describes the $\cN=1$ tensor multiplet \cite{Siegel})
and a set of chiral scalars. In our case, such a formulation may be obtained 
as follows. We note that the chiral superfields
\bea 
\phi_\pm = \Phi \pm  \ri \Psi
\label{pm}
\eea
transform under \eqref{4.8} as 
\bea
\d \f_\pm = \pm \ri \, \l \,\f_\pm~.
\label{polar}
\eea
We may introduce new local complex coordinates for the target space,  $\vf$ and $\c$, 
 that are defined as 
\begin{subequations}
\bea
\vf &=& \f_+ \f_- ~, \qquad \f_+ = \re^{  \c}~.
\label{phi-chi}
\eea 
Their U(1) transformation laws are respectively 
\bea
\d \vf &=& 0~, \qquad \d \c = \ri \l~. 
\label{phi-chi2}
\eea
\end{subequations}
It follows from these transformation laws that, 
in terms of the new coordinates introduced, the K\"ahler potential takes the form
\bea
K = K(\vf, \bar \vf, \c +\bar \c)~.
\label{Kp}
\eea
Next, the (anti) chiral variables $\c$ and $\bar \c$ can be dualised into 
a real linear superfield $G$ using the standard procedure \cite{Siegel,LR83}.
The resulting theory is described by a superfield Lagrangian ${\mathbb L}(\vf, \bar{\vf}, G) $.

It should be remarked that there is a considerable freedom in the choice of 
chiral superfields $\vf$ and $\c$ with the U(1) transformation laws \eqref{phi-chi2}. 
Instead of the variables \eqref{phi-chi}, one  equally well may deal with 
chiral superfields $\vf'$ and $\c'$ defined by 
\bea
\vf \to \vf' = f(\vf )~, \qquad \c \to \c' = \c + g (\vf )~,
\label{phi-chi3}
\eea
with $f(\vf )$ and $g(\vf )$ holomorphic functions. 
Implementing such a holomorphic field redefinition changes the K\"ahler potential
\eqref{Kp}, 
$K\to K'\equiv K'(\vf', \bar \vf', \c' +\bar \c') = K(\vf, \bar \vf, \c +\bar \c)$,
as well as leads to a modified chiral-tensor Lagrangian $\mathbb L' (\vf' ,\bar \vf' , G)$.

A large family of the U(1) invariant chiral $\s$-models 
discussed in this section
are in fact $\cN=2$ supersymmetric. 
For this to hold, the target space must be hyper K\"ahler \cite{A-GF}, 
and thus the corresponding Ricci tensor must vanish.\footnote{Any Ricci-flat K\"ahler manifold of real dimension four is 
hyper K\"ahler and vice versa, see e.g. \cite{Besse}.}
Using the condensed notation $\f^i = (\F, \J)$ and $\bar \f{}^{\bar i} = (\bar \F, \bar \J)$, 
the condition for Ricci-flatness is  the Monge-Amp\`ere equation 
(see, e.g., \cite{Besse})
\bea
\pa_k\pa_{\bar l} \ln \det \Big( \pa_i \pa_{\bar j} K (\f, \bar \f) \Big) =0~. 
\label{MA}
\eea
The Killing vector 
\bea
\k = \ri \Big( \frac{\pa}{\pa \c} - \frac{\pa}{\pa \bar \c}\Big)
\eea
is holomorphic with respect to the diagonal complex structure, but in general it is not 
triholomorphic.\footnote{It may be a Killing vector that rotates the two-sphere of complex structures, 
and thus  necessarily leaves one of the complex structures invariant \cite{HitchinKLR}.  
Such a Killing vector 
naturally originates in the case of $\cN=2$ supersymmetric $\s$-models on cotangent bundles
of K\"ahler manifolds \cite{GK1,GK2}. Moreover, the hyper K\"ahler target spaces of $\cN=2$ 
supersymmetric sigma models in four-dimensional anti-de Sitter space must possess such a Killing vector
\cite{BKsigma1,BKsigma2}. }
However, as demonstrated 
in \cite{HoweKLR}, 
if $\k$ is triholomorphic, 
one can associate with this hyper K\"ahler manifold an off-shell $\cN=2$ supersymmetric theory
describing a self-interacting $\cN=2$ tensor multiplet  \cite{LR83,KLR}.
The $\cN=2$ tensor multiplet theory is dynamically equivalent to the chiral $\s$-model constructed. 

It is possible to act in a reverse order. Let us start from the most general  
off-shell $\cN=2$ supersymmetric 
$\s$-model  describing a self-interacting $\cN=2$ tensor multiplet
\cite{LR83,KLR}. Formulated in $\cN=1$ superspace, the action is 
\begin{subequations} \label{4.15}
\bea
S_{\text{tensor}}= \int \rd^4x \rd^4 \theta \, {\mathbb L}(\vf, \bar{\vf}, G) ~,
\label{tensor}
\eea
where $\vf$ is a chiral scalar, $\bar D_\ad \vf=0$,
 and $G=\bar G$ is a real linear scalar, $\bar D^2 G=0$. 
The fact that the theory is $\cN=2$ supersymmetric means that the Lagrangian 
$\mathbb L$ has to obey the three-dimensional Laplace equation \cite{LR83},
\bea
\left( \frac{\pa^2 }{   \pa \vf \pa \bar \vf } +\frac{\pa^2 }{\pa G  \pa G} \right)
{\mathbb L}
=0~. 
\label{tensor-Laplace}
\eea
\end{subequations}
The theory \eqref{tensor} possesses a chiral formulation obtained by dualising $G$ 
into a chiral scalar $\c$ and its conjugate $\bar \c$ \cite{LR83}. 
The dual chiral action is 
\bea
S= \int \rd^4x \rd^4 \theta \,K(\vf, \bar \vf, \c +\bar \c)~,
\label{tensor2}
\eea
and it is manifestly U(1) invariant. 
What is special about this particular complex parametrisation of the hyper K\"ahler target space 
is the unimodularity condition   \cite{LR83}
\bea
\frac{\pa^2 K}{\pa \c \pa \bar \c} \frac{\pa^2 K}{\pa \vf \pa \bar \vf} 
- \frac{\pa^2 K}{\pa \c \pa \bar \vf} \frac{\pa^2 K}{\pa \vf \pa \bar \c}  =1~,
\label{MA2}
\eea
which is a stronger version of  eq. \eqref{MA}.
Next, this $\s$-model \eqref{tensor2} can be reformulated in terms of the chiral superfields $\F$ and $\J$ 
defined according to eqs. \eqref{pm} and \eqref{polar}.
The resulting nonlinear $\s$-model $S_{\text{chiral}}$, eq. \eqref{4.4}, 
can equivalently be realised as a U(1) duality  invariant CCL $\s$-model.

As a result,  we have established a  correspondence between 
self-interacting $\cN=2$ supersymmetric tensor multiplet models 
and certain U(1) duality invariant CCS $\s$-models \eqref{2.1} with hidden $\cN=2$ supersymmetry. 
The condition \eqref{MA2} should be imposed in order 
to fix the freedom \eqref{phi-chi3}
when going from the CCS formulation to the tensor one. 

As noted in section 2, if the original Lagrangian 
$L(\Phi, \bar{\Phi}, \S, \bar{\S})$ can be expressed in the form $L(\o, \bar{\o}),$ 
then $L$ obeys two differential equations \eqref{3.12}. 
In the chiral formulation, the first of these equations,  \eqref{3.12a}, turns into 
\bea
\frac{\partial K}{\partial \F} \F + \frac{\partial K}{\partial \Psi} \J
= \frac{\partial K}{\partial \bar \F} \bar \F + \frac{\partial K}{\partial\bar  \Psi} \bar \J~.
\eea
This is the condition for the invariance of 
$K(\Phi, \bar{\Phi}, \Psi, \bar{\Psi})$  under additional U(1) transformations
\bea
\d \Phi = \ri \a \F, \quad \d \Psi =  \ri \a \J, \qquad \a \in {\mathbb R}~.
\label{4.10}
\eea
Thus the isometry group of the $\sigma$-model target space includes
the group $\rm U (1) \times \rm U (1)$ of transformations \eqref{4.8} and 
\eqref{4.10}. These symmetries imply that the K\"ahler potential can be represented 
as a function of two real variables
\bea
K(\Phi, \bar{\Phi}, \Psi, \bar{\Psi})= K(\o_+, \o_-)~,
\label{4.16}
\eea
where 
\be 
\o_{\pm} = \bar{\Phi} \Phi + \bar{\Psi} \Psi \pm \ri \, (\bar{\Phi} \Psi- \bar{\Psi} \Phi  )~.
\label{omegaplusminus}
\ee
The simplest way to show this is to switch from the chiral scalars $\F$ and $\J$ to  
the new chiral variables \eqref{pm}. The Abelian transformations \eqref{4.8} and 
\eqref{4.10} can be realised on these variables as 
\begin{subequations}\label{4.18}
\bea
\d \f_+ &=& \ri \l_+ \f_+~, \quad \d \f_- =0~, 
 \\
\d \f_+ &=&0~, \qquad  \quad \d \f_- = \ri \l_- \f_- ~,
\eea
\end{subequations}
with real parameters $ \l_\pm \in {\mathbb R}$. 
Since the K\"ahler potential must be invariant under the transformations \eqref{4.18}, 
we conclude that $K= K(\f_+ \bar \f_+, \f_- \bar \f_-)$.
It remains to note that the real variables \eqref{omegaplusminus}
can be factorized in the form 
\be
\o_+ = \phi_+ \bar{\phi}_+ ~, \quad \o_- = \phi_- \bar{\phi}_-~.
\ee
In particular,  they are invariant under the transformations \eqref{4.18}.

So far we have only derived the implications of \eqref{3.12a} 
in the dual chiral representation 
of duality invariant CCL models (\ref{2.1}).  
The counterpart of  equation (\ref{scale}) in the dual chiral formulation is
\be
0 = \Phi \bar{\Psi} - \bar{\Phi} \Psi 
+ \frac{\partial K}{\partial \Phi} \frac{\partial K}{\partial \bar{\Psi}} 
-  \frac{\partial K}{\partial \bar{\Phi}} \frac{\partial K}{\partial \Psi}~ .
\label{dualSS}
\ee
By comparison with (\ref{SS}), this looks like a condition for duality invariance 
in the dual $\Phi$-$\Psi$ sector. Its origin is the formal symmetry 
$\delta \Phi = \ri  \l \, \bar{\S}$ and $ \d \bar{\S} = - \ri  \l \, \Phi$,
in the $\Phi$-$\S$ sector, just as the duality equation (\ref{SS}) 
in the $\Phi$-$\S$ sector can be considered as being a consequence 
of U(1) symmetry (\ref{SO2}) in the chiral $\Phi$-$\Psi$ sector via a Legendre transform.

In terms of  the real variables \eqref{omegaplusminus}, 
 the invariance condition  (\ref{dualSS}) for the K\"ahler potential 
\eqref{4.16}  can be expressed in the form
\be
0 = \o_+ - \o_- - 4 \, \o_+ \left( \frac{\partial K}{\partial \o_+} \right)^2 
+ 4 \,\o_- \left( \frac{\partial K}{\partial \o_-} \right)^2~.
\label{pmdual}
\ee
This is structurally of the same form as the duality invariance condition (\ref{universal}) for nonlinear electrodynamics; however, here $\o_{\pm}$ are real variables, whereas in (\ref{universal}), $\o$ is a complex variable.
A general solution of equation \eqref{pmdual} has the form 
\bea
K(\o_+, \o_-) &=&  \hf \bar \f_+ \f_+ + \hf \bar \f_- \f_-
+\hf \sum_{m, n=1}^{\infty} C_{m,n} (\bar \f_+ \f_+)^m (\bar \f_- \f_-)^n~,
\eea
with $C_{m,n}$ real coefficients. 
Equation  \eqref{pmdual}  proves to determine all 
the coefficients $C_{m,n}$ with $m\neq n$ in terms of those with $m=n$,
with the latter being completely arbitrary. As a result, the general solution of \eqref{pmdual} 
involves an arbitrary real function of a real argument, $f(x) = \sum C_{n,n} x^n$.
The situation is completely analogous to that in self-dual nonlinear electrodynamics, 
see \cite{KT2} for  a review. 

It is interesting that equation  \eqref{pmdual} does not allow higher-order 
K\"ahler-like contributions to the K\"ahler potential 
of the form $(\bar \f_+ \f_+)^n$, with $n>1$ (and similarly in the $\f_-$ sector).
This may be interpreted as 
a non-renormalisation theorem that keeps the  tree-level kinetic term
$ \o_+ = \bar \f_+ \f_+$ for  the field $\f_+$ and $\bar \f_+$
protected against  ``quantum'' nonlinear corrections.

\section{Dual Born-Infeld type solutions} \label{section5}
\setcounter{equation}{0}

As discussed in section 3, the Lagrangian
\be 
L_{\rm BI}(\Phi, \bar{\Phi}, \S, \bar{\S}) =  \frac{1}{g} \, (1 - \D^{\frac12}) 
\ee
with 
\be
\D = 1 - 2 g ( \bar{\Phi} \Phi - \bar{\S} \S) - g^2 (\Phi \S + \bar{\Phi} \bar{\S})^2
\ee
is duality invariant.  When expressed in terms of the the complex variables $\o$ and $\bar{\o} $ defined in (\ref{omega}), the functional form of this Lagrangian is the same as that for the famous Born-Infeld action for nonlinear electrodynamics.

Dualisation of the complex linear superfields $\S$ in favour of chiral scalar superfields $ \Psi$ is via the the Legendre transform
\be
 L_{\rm D} (\Phi, \bar{\Phi}, \Psi, \bar{\Psi}) 
 = \frac{1}{g} (1 - \D^{\frac12})  + \Psi \S + \bar{\Psi} \bar{\S}~.
 \label{SD}
 \ee
Eliminating  $\S$  and $\bar{\S}$ by their equations of motion, we obtain the implicit equation
\be
\left(
\begin{array}{c}
\S  \\  \bar{\S}  
\end{array}
\right)  = \Delta^{\frac12} \, (1 - A)^{-1} \, \left(
\begin{array}{c }
\bar{\Psi} \\  \Psi  
\end{array}
\right)~,
\ee
where 
\be
A = g \left(
\begin{array}{cc}
 \bar{\Phi} \Phi &  \quad  \bar{\Phi} \bar{\Phi}\\
  \Phi \Phi & \quad  \Phi \bar{\Phi}
\end{array}
\right)~. 
\ee
Substituting into (\ref{SD}), the dual Lagrangian  is 
\be
L_{\rm D} (\Phi, \bar{\Phi}, \Psi, \bar{\Psi})  
=   1 -  (1 - 2 g \, \bar{\Phi} \Phi)^{\frac12} (1 - g  \a + g^2  \b^2)^{-\frac12} \, (1 - g  \g ) 
\label{SD1} 
\ee
with
\begin{eqnarray}
\a & = &
\left(
\begin{array}{cc}
\Psi \, , &  \bar{\Psi }  
\end{array}
\right) \, (1 - A)^{-2}  \, \left(
\begin{array}{c}
\bar{\Psi }  \\ 
\Psi  
\end{array}
\right), 
\quad
 \b = \left(
\begin{array}{cc}
\Phi  \, , &  \bar{\Phi }  
\end{array}
\right) \, (1 - A)^{-1}  \, \left(
\begin{array}{c}
\bar{\Psi }  \\ 
\Psi  
\end{array}
\right), \\
 \g &= & \left(
\begin{array}{cc}
\Psi \, , &  \bar{\Psi }  
\end{array}
\right) \, (1 - A)^{-1}  \, \left(
\begin{array}{c}
\bar{\Psi }  \\ 
\Psi  
\end{array}
\right).
\end{eqnarray}

Representing the matrix $A$ in the form
\be
A = g \left(
\begin{array}{c}
\bar{\Phi}  \\  \Phi  
\end{array}
\right) \,
\left(
\begin{array}{cc}
\Phi \, , &  \bar{\Phi}  
\end{array}
\right)
\ee
allows $\a,$ $\b$ and $\gamma$ to be expressed as
\bea
\a &=& 2 \bar{\Psi} \Psi + 2 (1 - 2g \, \bar{\Phi} \Phi)^{-2} \, (\bar{\Psi} \Phi + \bar{\Phi} \Psi)^2 \\
\b &=& (1 - 2 g \, \bar{\Phi} \Phi)^{-1} \, (\bar{\Psi} \Phi + \bar{\Phi} \Psi) \\
\gamma &=& 2 \bar{\Psi} \Psi + (1 - 2 g \, \bar{\Phi} \Phi)^{-1} \, (\bar{\Psi} \Phi + \bar{\Phi} \Psi)^2~.
\eea
Substituting into (\ref{SD1}) yields
\be
L_{\rm D} (\Phi, \bar{\Phi}, \Psi, \bar{\Psi}) =  \frac{1}{g} (  1 - \D_{\rm D}^{\, \, \frac12} )
\ee 
where
\be 
\D_{\rm D}  = 1 - 2g \, (\bar{\Phi} \Phi + \bar{\Psi} \Psi )  - g^2 (\bar{\Psi} \Phi - \bar{\Phi} \Psi )^2~.
\ee
This can be expressed as in terms of the real variables $\o_{\pm},$ defined in (\ref{omegaplusminus}), as
\be 
\D_{\rm D} = 1 - g \, (\o_+ + \o_-) - \frac{g^2}{4} (\o_+ - \o_-)^2~,
\ee
which exhibits a Born-Infeld type functional form for the dual chiral Lagrangian. By construction, this dual Lagrangian  is  a solution of  (\ref{pmdual}).

Using similar techniques for  the variant Born-Infeld type Lagrangian 
 involving $\widetilde{\D}$  defined  in (\ref{Delta1}), the dual Lagrangian is
\be
 \widetilde{L}_{\rm D} =  \frac{1}{g} (1 - \widetilde{\D}_{\rm D}^{\, \, \frac12} )~,
\ee
with
\be
\widetilde{\D}_{\rm D} = 1 - 2g \, (  \bar{\Phi} \Phi +  \bar{\Psi} \Psi)~.
\ee

\section{Conclusion} 
\setcounter{equation}{0}

Recently, there has been a revival of interest in the duality invariant dynamical systems
of Abelian vector fields, see \cite{BN,CKR,BCFKR,RT} and references therein.  
This interest was mainly inspired by the desire to achieve a better understanding
of   the UV properties of extended supergravity theories. The recent studies have concentrated,
in particular,  
on the following two problems: (i) consistent deformations of duality invariant systems
 \cite{BN,CKR,BCFKR}; and (ii) the fate of duality invariance in quantum theory \cite{RT}.
Duality invariant CCL $\s$-models may shed some light on both of these issues. 

Let us first briefly discuss the problem (i). 
In the case of nonlinear electrodynamics, the condition of 
duality invariance \eqref{universal} is a nonlinear differential equation on the 
Lagrangian. This means that the problem of consistent deformations  
of duality invariant theories is rather nontrivial.
In order to develop a systematic procedure to generate duality invariant theories,
the authors of \cite{BN,CKR} put forward the so-called ``twisted self-duality constraint'' 
and provided simple perturbative applications of this scheme.  
However, it has been demonstrated by Ivanov and Zupnik  \cite{IZ3} 
that the non-supersymmetric constructions of  
\cite{BN,CKR} naturally originate within the more general approach 
developed a decade earlier in \cite{IZ1,IZ2}. Specifically, the twisted self-duality constraint 
corresponds to an equation of motion in the approach of \cite{IZ1,IZ2}. 
The approach of \cite{IZ1,IZ2} has also been extended \cite{K2013,ILZ} 
to the cases of duality invariant $\cN=1$ and $\cN=2$ (locally) supersymmetric 
vector multiplet models \cite{KT1,KT2,KMcC,K_N=2}. The main idea of the generating formalism 
\cite{IZ1,IZ2} consists in reformulating the U(1) duality invariant models by introducing auxiliary 
variables in such a way that the self-interaction is manifestly U(1) invariant.
The original theory is obtained by integrating out the auxiliary variables.
Since any choice of U(1)  invariant  self-interaction proves to lead to a U(1) duality invariant 
model, the Ivanov-Zupnik approach \cite{IZ1,IZ2} is an efficient scheme to generate 
duality invariant systems. 

In the case of U(1) duality invariant CCL $\s$-models, there is no need to introduce 
auxiliary variables as a mechanism to generate such dynamical systems. 
As we demonstrated in section 4, the dual chiral representation plays the role of such a generating formalism. 
 
Regarding the problem (ii) raised e.g. in \cite{RT}, the main issue is that duality invariance
is not a manifest symmetry of the action. As a result, the precise realisation of this symmetry 
at the S-matrix level, or in the framework of the effective action, 
 requires an additional definition. This issue is nontrivial, for instance,  in the case of
the duality invariant models for nonlinear electrodynamics 
(barring the non-renormalizability of such models). 
In the case of duality invariant CCL $\s$-models, we have
a natural way out. These theories possess the dual chiral formulation in which 
U(1) duality symmetry turns into a manifest  U(1) symmetry. 
Switching to the dual chiral formulation requires us to make use of 
the superfield Legendre transformation described in section \ref{section4}, 
and the latter can naturally be implemented within the path integral. 

We see that there is a conceptual difference between the two families 
of U(1) duality invariant theories: (i) models for nonlinear electrodynamics 
and (ii) CCL $\s$-models. This difference is that only for the latter family do we have 
the ability to realise continuous duality symmetries as manifest U(1) symmetries 
in a dual formulation of a theory.
 Still, there is a way to relate these two types of theories.
 Specifically, we can start with a four-dimensional duality invariant model for nonlinear electrodynamics 
 and dimensionally reduce it to three dimensions 
 resulting in a duality invariant theory involving a vector field and 
 a scalar field.\footnote{This is a specific example  of more general duality invariant  systems 
 of $(p - 1)$ forms and $(d-p-1)$ forms in $d$ space-time dimensions, as discussed in section 8 of \cite{KT2}).} 
 By Legendre transformation, this theory can be cast purely in terms of scalar fields 
 or purely in terms of vector fields, each involving a manifest U(1) symmetry that has its origin in the duality invariance of the original theory.

In this paper, our discussion was restricted to $\cN=1$ SUSY in four dimensions. 
Plain dimensional reduction leads to 
analogous results for $\cN=2$ SUSY in three dimensions and 
$\cN=(2,2)$ SUSY in two dimensions. 
In fact, it is well known that the two-dimensional 
$\cN=(2,2)$ Poincar\'e supersymmetry allows the existence 
of new superfield types that are impossible in higher dimensions:
 twisted chiral \cite{GHR} and   semi-chiral \cite{BLR}. 
In terms of such supermultiplets one may define new families of 
duality invariant $\s$-models.

We have also restricted our discussion to the case of U(1) duality invariant CCL $\s$-models with a single chiral multiplet and a single complex linear multiplet. 
Inclusion of $n>1$ chiral -- complex linear doublets  is expected 
to allow non-Abelian duality groups.\footnote{CCL $\s$-models with non-Abelian duality groups
may be obtained by generalising the construction of the general duality invariant  systems 
 of $(p - 1)$ forms and $(d-p-1)$ forms in $d$ space-time dimensions, see section 8 of \cite{KT2}.
Alternatively, one may start from a chiral $\s$-model with a manifest non-Abelian symmetry group
and then  switch to  the dual chiral-complex linear formulation.}
It is also of interest to consider gauging the isometries in the purely chiral chiral theory and 
investigate the consequences in the dual chiral -- complex linear theory. 
\\

\noindent
{\bf Acknowledgement:}
The work of SK is supported in part by the Australian Research Council. 

\appendix

\section{The duality equation} 
\setcounter{equation}{0}

Here, we adapt standard arguments (as reviewed in \cite{KT2}) from nonlinear electrodynamics to derive  
an integrability condition associated with the consistency of the definition of $L'(X')$ 
via equations (\ref{2.3a}) and \eqref{2.3b} that requires $\k = \l.$ We then derive  
the requirement (\ref{SS}) for duality invariance  of nonlinear $\s$-models \eqref{2.1}
 under the duality rotations (\ref{2.3}). As in earlier sections, we use the notation $X$ 
 for the argument $\Phi, \bar{\Phi}, \S, \bar{\S}.$

 In infinitesimal form, the U(1) duality rotations (\ref{2.3}) preserving the constraints and the equations of motion are
 \bea
 \frac{\partial L'(X')}{\partial \S'} &=& \frac{\partial L(X)}{\partial \S} - \l \, \Phi, \quad \Phi' 
 = \Phi + \l \, \frac{\partial L(X)}{\partial \S}~,
 \label{inf1} 
 \\
  \frac{\partial L'(X')}{\partial \Phi'} &=& \frac{\partial L(X)}{\partial \Phi}- \k \,  \S, \quad \S' 
  = \S + \k \, \frac{\partial L(X)}{\partial \Phi}~.
 \label{inf2}
 \eea
  
Using the notation
 \be
 \d X \, \frac{\partial L(X)}{\partial X} \equiv \d \Phi \, \frac{\partial L(X)}{\partial \Phi} 
 + \d \bar{\Phi} \, \frac{\partial L(X)}{\partial \bar{\Phi} } 
 +  \d \S \, \frac{\partial L(X)}{\partial \S} 
 + \d \bar{\S} \, \frac{\partial L(X)}{\partial \bar{\S}}~,
 \ee
 then to first order in $\d X,$
 \be
 L'(X') = L(X) + \D L(X) + \d X \, \frac{\partial L'(X)}{\partial X}~,
 \ee
 where 
 \be 
 \D L(X) = L'(X) - L(X)~.    
 \label{DL}                  
 \ee
 Using the chain rule to convert derivatives with respect to $\S'$ 
 into derivatives with respect to $\Phi, \S, \bar{\Phi}$ and $\bar{\S},$ and retaining only 
 terms linear in the infinitesimal parameters $\k$ and $\l,$
 the first equation in (\ref{inf1}) becomes
 \bea
- \l \, \Phi &=&  \frac{\partial}{\partial \S} \left( \D L(X) + \d X \frac{\partial L(X)}{\partial X} \right) - \k \, \frac{\partial^2 L(X)}{ \partial \S \, \partial \Phi } \,\frac{\partial L(X)}{\partial\S} \nonumber \\
 &- & \k \, \frac{\partial^2 L(X)}{\partial \S \, \partial \bar{\Phi}  } \,\frac{\partial L(X)}{\partial \bar{\S} } 
 -  \l \,  \frac{\partial^2 L(X)}{\partial\S^2} \,\frac{\partial L(X)}{\partial \Phi} - \l  \, \frac{\partial^2 L(X)}{\partial \S \, \partial \bar{\S} } \,\frac{\partial L(X)}{\partial \bar{\Phi}} ~. 
 \label{deriv}
 \eea
The left hand side of this relation can be expressed as the derivative 
$\frac{\partial}{\partial \S} (- \l  \Phi \S),$ so consistency requires that the right hand side can be expressed as a derivative with respect to $\S.$ This is only so if $\k = \l,$ in which case (\ref{deriv}) is 
\be
0 = \frac{\partial}{\partial \S} \left( \D L(X) +  \l \, \Phi \, \S +\d X \frac{\partial L(X)}{\partial X}  - \l \, \frac{\partial L(X)}{\partial \Phi} \, \frac{\partial L(X)}{\partial \S} - \l \,  \frac{\partial L(X)}{\partial \bar{ \Phi}} \, \frac{\partial L(X)}{\partial \bar{\S}} \right)~.
\ee
Integrating and requiring reality of $L(X)$ yields
\be
\D L (X) = - \l \, \Phi \, \S  - \l \, \bar{\Phi} \, \bar{\S} - \l \, \frac{\partial L(X)}{\partial \Phi} \, \frac{\partial L(X)}{\partial \S} -  \l \, \frac{\partial L(X)}{\partial \bar{ \Phi}} \, \frac{\partial L(X)}{\partial \bar{\S}}~.
\label{deltaL1}
\ee

Imposing the requirement (\ref{dualityinv}) for duality invariance means that $\D L(X) $ defined in (\ref{DL}) vanishes. Equation (\ref{deltaL1}) then yields the condition (\ref{SS}) for duality invariance of the Lagrangian $L(X).$

\begin{footnotesize}

\end{footnotesize}

\end{document}